\documentclass{aa}
\usepackage{graphics}
\begin{document}

\thesaurus{?? (04.01.1; 04.03.1; 01.16.1; 04.19.1; 01.19.2)}

\title{The CDS information hub}

\subtitle{On--line services and links at the Centre de Donn\'ees 
astronomiques de Strasbourg}

\author{Fran\c{c}oise Genova, Daniel Egret, Olivier Bienaym\'e, Fran\c{c}ois 
Bonnarel, 
Pascal Dubois, Pierre Fernique, 
G\'erard Jasniewicz\thanks{Groupe de Recherche en
Astronomie et Astrophysique du Languedoc (GRAAL), Montpellier}, 
Soizick Lesteven, 
Richard Monier, Fran\c{c}ois Ochsenbein, Marc Wenger}

\institute{CDS, Observatoire astronomique de Strasbourg, UMR 7550, 11
rue de l'Universit\'e, F--67000 Strasbourg, France \\
email: question@simbad.u-strasbg.fr}

\date{Received \today/ Accepted}

\authorrunning{Fran\c{c}oise Genova et al.}

\offprints{Fran\c{c}oise Genova}

\maketitle

\begin{abstract}

The {\it Centre de Donn\'ees astronomiques de Strasbourg} (CDS)
provides homogeneous access to heterogeneous 
information of various origins: information about 
astronomical objects in {\sc Simbad};
catalogs and observation logs in {\sc VizieR} and in the
catalogue service; reference images and overlays 
in {\sc Aladin}; nomenclature in the {\it
Dictionary of Nomenclature}; Yellow Page services; the AstroGLU resource
discovery tool; mirror copies of other reference services; and
documentation. With the implementation of links between
the CDS services, and with other on--line reference information, CDS
has become a major hub in the rapidly evolving world of information
retrieval in astronomy, developing efficient tools to help astronomers
to navigate in the world-wide `Virtual
Observatory' under construction, from
data in the observatory archives to results published in journals.

The WWW interface to the CDS services is available at:

http://cdsweb.u-strasbg.fr/

\keywords{Astronomical data bases: miscellaneous -- Catalogs
-- Publications, bibliography -- Surveys --  Standards}
\end{abstract} 

\section{Introduction}

The {\it Centre de Donn\'ees astronomiques de Strasbourg} (CDS) was
founded in 1972 as the {\it Centre de Donn\'ees Stellaires}, and
installed in Strasbourg as the result of an agreement
between the INAG ({\it Institut National d'Astronomie et de
G\'eophysique}), now INSU ({\it Institut National des Sciences de
l'Univers}), and {\it Universit\'e Louis Pasteur}. This was
the outcome from the
very prospective vision of Jean Delhaye, then Director of INAG, who 
anticipated the importance of computer-readable
data. It was decided to found CDS as a part of the French research system,
as a Data Center serving the international astronomical scientific
community.

The objectives of CDS at its creation could be summarized as follows :

\begin{itemize}
\item
collect `useful' data about astronomical sources, in electronic form;
\item
improve them by critical evaluation and combination;
\item
distribute the results to the international community;
\item
conduct research programs using the data collected.
\end{itemize}

Insertion of CDS into a research institution, 
the {\it Observatoire astronomique de Strasbourg}, 
helps to maintain direct
contacts with the evolution of astronomy, and with researchers' actual needs.

At the beginning, CDS was dealing with stellar data, 
aiming at the study of the galactic structure. In 1983, 
it was decided that {\sc Simbad}, one of the two important
CDS services at that time, would also deal with other galactic and
extragalactic objets -- i.e., with all astronomical objects outside the
Solar System. The CDS's name was changed to {\it Centre
de Donn\'ees astronomiques de Strasbourg}, thus preserving the acronym
which was already well known.

In recent years, research activities in astronomy have 
evolved significantly, with the very rapid development 
of on-line information at all levels, from
observatory archives to results published in journals.
The challenge is now to deal with {\it information} more than 
with {\it
data}, which includes data, but also know-how about data,
technical information about instruments, published results,
compilations, etc.
 
The CDS goals can now be summarized 
as {\bf collect, homogenize, distribute, and preserve astronomical
information for the scientific use of the whole astronomical 
community}. This `mission
statement' still
contains all the drivers from the early CDS charter: 
dealing with electronic data,
taking up an international role, developing expertise on astronomical
data, having research as a goal. 

All on-line CDS services can be accessed from the CDS home page
on the World-Wide Web\footnote{http://cdsweb.u-strasbg.fr/}.

\section{The present context of CDS activities}

The ground- and space-based observatories, the sky surveys,
and the deep field observations, produce 
large amounts of data, obtained at different wavelengths with different
techniques. To understand the physical phenomena at work in objects,
astronomers need to access this wealth of data, to understand
their meaning (error bars, etc.), and to combine the use of data from different
origins, especially with  the
development of {\it panchromatic} astronomy. At all steps of an
astronomer's work, it is thus more and more necessary to
re-use data obtained by others and to take into account
previous  results, often from other fields of astronomy. 

Astronomers and funding Agencies are now well aware of
the necessity of preserving and diffusing data and results. This leads to
several types of developments:

\begin{itemize}

\item the data producing teams, which have the knowledge of instruments
and observation techniques, have to preserve all or part of their data,
in a form usable by all astronomers -- in the majority of
recent large projects, 
the building of the {\it observatory archive} is now considered as one of
the project missions;

\item in some domains, a {\it specialized center} provides access to and
information about data in a given subfield of astronomy -- this is the
case in particular of the
disciplinary NASA Centers, HEASARC\footnote{http://heasarc.gsfc.nasa.gov/}
 for High Energy astrophysics, IRSA\footnote{http://irsa.ipac.caltech.edu/}
at 
IPAC\footnote{http://www.ipac.caltech.edu/} for Infrared data, 
and MAST\footnote{http://archive.stsci.edu/mast.html}
at the Space Telescope Science Institute\footnote{http://www.stsci.edu/},
for optical and UV data;

\item as a Data Center, the role of CDS 
is to bridge the gap between the specialized approach of the
scientific teams, and the general approach of the community of
researchers. In the present context, with the very rapidly growing
number of on-line services of interest to astronomers, this
means in practical terms the definition, development, and
distribution of tools for retrieving useful information among
the vast array of possible sources (Sections 3, 5).

\end{itemize}

In addition, journals are recent, but very active actors in
the on-line distribution of information in astronomy. Electronic
publication has made rapid progress
and many journals now have on-line
versions, which often display external links. Moreover, NASA {\it
Astrophysics Data System}\footnote{http://ads.harvard.edu/, with several
mirror sites, including one at CDS at http://cdsads.u-strasbg.fr/}
(ADS, Kurtz et al. \cite{ads}) has become a major reference tool for
astronomers. Several aspects of the ADS are described in a set of
companion papers.
The bibliographic astronomy network, and the CDS role in this
networking, are described in more detail in Section 5.2.

On the other hand, computers and
networks evolve very rapidly: the high rate of increase in the volume 
of data and results,
the irruption of the Internet and World Wide Web,
the widespread usage of graphical interfaces, the lower and lower cost of
information storage, completely changed the technical context of the CDS
activities in the last few years.

\section{The CDS activities}

CDS activity has different aspects. Some are directly visible to
the users, whereas others, though fundamental for maintaining the CDS
expertise and role, may be less conspicuous.

The most perceptible activity of CDS is certainly the development,
maintenance and on-line diffusion of reference, value-added
databases and services, such as
{\sc Simbad}, {\sc VizieR}, {\sc Aladin},
the {\it Dictionary of Nomenclature of
astronomical objects outside the solar system} (Lortet et al.
\cite{dic}), the AstroGLU discovery tools (Egret et al. 
\cite{astroglu}), etc. 
The CDS services are described in more detail in Section 4,
and in the set of companion papers by Wenger et al. (\cite{simbad}),
Ochsenbein et al. (\cite{vizier}), and Bonnarel et al. (\cite{aladin}).

From the point of view of contents, CDS deals with
selected information: raw observational data are generally not available
at CDS, but rather upper level data such as observation logs,
catalogues, results, etc.
This `reference' information is then documented, organized, and made
accessible in the CDS services.

In addition, in order to cope with the congestion of
inter-continental connections, CDS has developed an active
policy of mirror copy implementation. 
The redundant availability of data on several sites is also
important to ensure data security.
Mirror copies of {\sc Simbad} and ADS are installed in CfA and CDS
respectively (Eichhorn et al. \cite{mirror}). 
Mirror copies of {\sc VizieR} are installed 
at NASA ADC\footnote{http://adc.gsfc.nasa.gov/} 
and NAOJ ADAC\footnote{http://adac.mtk.nao.ac.jp/}
(which also hosts a copy of the {\it Dictionary of
Nomenclature}), another one is foreseen at the Indian Data Center 
(IUCAA\footnote{http://www.iucaa.ernet.in/}, Pune). 
CDS hosts mirror copies of several electronic journals
and of CFHT documentation.

Less visible from the users, but an important field for
networking international partnership, 
CDS develops generic tools and distributes them to other information
providers: for example,
the GLU ({\it G\'en\'erateur
de Liens Uniformes}), for maintaining links to
distributed heterogeneous databases (Fernique et al.
\cite{glu}, Section 5.1), or
the {\sc Simbad} client/server package, which allows archive services and the
ADS to use {\sc Simbad} as a name resolver.

CDS is also active in the development of exchange standards, 
such as the {\it bibcode}, first defined by
NED\footnote{http://nedwww.ipac.caltech.edu/}
and the CDS (Schmitz et al. \cite{bibcode}), 
and now widely used by the ADS and the on-line
journals, or the standard description of tables, defined by CDS and
shared with the other data centers and the journals (Ochsenbein et al.
\cite{vizier}).
CDS collaboration with journals is described in Section~5.2.

CDS expertise in the domain of astronomical data is also useful for
projects. Strasbourg Observatory has been deeply involved in
construction of the input and result catalogues of Hipparcos and
Tycho (\cite{hip}),
and {\sc Simbad} has been used as a basis for ROSAT identification tools.
At present, the XMM Survey Science
Center\footnote{http://xmmssc-www.star.le.ac.uk/}
relies on the CDS services to build a
cross-identification database for the X-ray sources observed by
the satellite. The CDS participation to the DENIS and
TERAPIX/MEGAPRIME surveys is described in Section 5.4.

Finally, to keep up with the pace of technical
evolution, CDS has to develop a resolute activity in the domain of
technological and methodological watch, and to undertake
Research and Technology (R\&T) actions to assess new techniques.
The GLU development (Fernique et al. \cite{glu}, 
Section 5.1.), and the {\sc Aladin} Java
interface (Fernique \& Bonnarel, \cite{java}),
are examples of R\&T actions which came out as operational
services. More recently, one can cite the {\it ESO/CDS Data Mining Project}
(Ortiz et al. \cite{datamining}), 
or the assessment of commercial object oriented
database systems (Wenger et al. \cite{oodb}) (Section 5.4).

In practice, the main challenge in the CDS activity is to
constantly tune the contents and the services to the rapid scientific
and technical evolution, to be able to deal with ever increasing volumes
of information, and to be ready to respond to the new projects and
to the evolution of the policy of the national and international
Agencies. The guidelines in prioritizing the tasks are several:
offer the best service to the users, ensure the quality of
contents, and make the best of technological innovation. This
implies that new developments are begun neither too early, to rely on
techniques which are as secure as possible, nor too late, to offer the
best possible services and improve the functionalities -- hence
the importance of technical watch. A balance has constantly to
be kept, between very long term activities, on time scales of
several ten years, to build up the contents; the development
and maintenance of database systems and user interfaces, on
time scales of a few months to a few years; R\&T activities, on
similar times scales ; and operational constraints
on a day-to-day basis. Hence the
importance of careful strategy definition and  activity
scheduling. 

\section{The main CDS services}


\begin{figure}
\resizebox{\hsize}{!}{\includegraphics{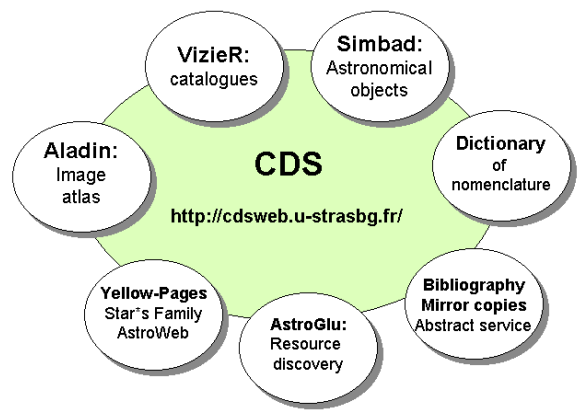}}
\caption{The main CDS on-line services.}
\label{cdsorg}
\end{figure}


\begin{figure}
\resizebox{\hsize}{!}{\includegraphics{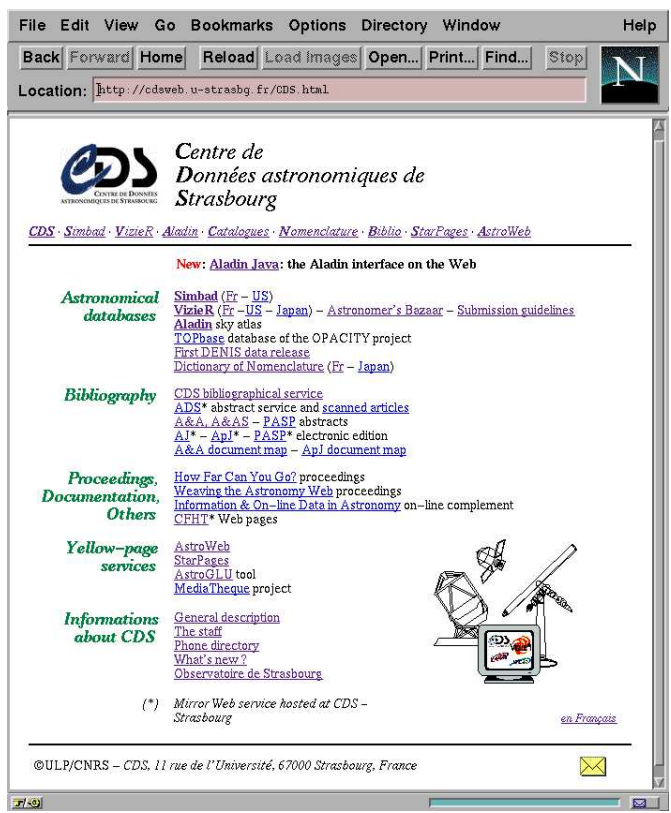}}
\caption{CDS home page on the World-Wide Web.}
\label{cdshome}
\end{figure}

A diagram of the main CDS services is shown in
Fig.~\ref{cdsorg}, and their list is given in the CDS Home
Page (Figure~\ref{cdshome}).

The first two main products of CDS have been
the collection of information about astronomical objects from published
papers and reference catalogues in {\sc Simbad}, and the collection,
documentation, long term storage 
and distribution of catalogues in the catalogue service, with
the recent addition of the catalogue browser functionality of
the {\sc VizieR} service. 
More recently, the {\sc Aladin} project has permitted us
to construct an  image server, and a comprehensive tool to overlay the
information from {\sc Simbad}, {\sc VizieR}, and from other sources such
as NED or data archives,  on digitized images of the sky. These services
are described  in companion papers
(Wenger et al. \cite{simbad}, Ochsenbein et al. \cite{vizier}, Bonnarel et
al. \cite{aladin}).

The CDS also has the
responsibility of the {\it Dictionary of Nomenclature} (Lortet et al.
\cite{dic}),
hosts bibliographical information, with
mirror copies of the ADS and of the {\it Astrophysical Journal}, the {\it
Astronomical Journal}, and the {\it Publications of the Astronomical Society
of the Pacific}, and develops
bibliographic information retrieval tools (Poin\c{c}ot et al.
\cite{kohonen}). It maintains the AstroGLU information discovery tool
(Egret et al. \cite{astroglu}), and hosts two Yellow Page services: 
AstroWeb (Jackson et al. \cite{astroweb}), and
the {\it Star*s Family of Astronomy and Related Resources}
(Heck \cite{starpages}). CDS is the French IUE National host,
the host of a copy of the
CFHT user documentation and 
of the unpublished data on variable stars of IAU Commission 27.

The main evolution of the CDS in recent years, is the rapid
development of the {\it World Wide Web} access to the 
services\footnote{http://cdsweb.u-strasbg.fr/}. Before
1996, the only CDS service available on the Web, besides documentation,
was the access to the catalogues and tables (via ftp) and to the 
on-line abstracts of
the journal {\it Astronomy and Astrophysics}. {\sc VizieR} was released in
February 1996, the first Web version of {\sc Simbad} in November 1996, and the
Web access to {\sc Aladin} in November 1998 (Previewer) and February 1999
({\sc Aladin} Java).  Now all the services are accessible from the CDS Home
Page (Figure 2).

The usage of the CDS services has been continuously
increasing, with over 6,000 queries submitted to the CDS
services and their mirror copies every day,
and over 25,000 hits per day on the Web pages (November 1999).

\section {Major evolution trends}

With the development of the World Wide Web, building links 
between heterogeneous, distributed information,
has been an important evolution trend recently, and it is also an
important 
topic for international partnership among service providers. 

\subsection{Building links between heterogeneous services}

Historically, the CDS services had developed separately, with different
contents, functionalities, database management systems and user
interfaces. The World Wide Web opened the possibility to increase the
synergy between the services, by building links allowing the users to
navigate in a transparent way.
Maintenance is a major challenge however, as soon as
one tries to build links between distributed, heterogeneous services:
any change in the service address, or in the query syntax, breaks links.
This is particularly difficult when `anchors' (links in HTML syntax) are
hard-coded in the HTML pages. CDS
has solved this problem by developing
the {\it G\'en\'erateur de Liens Uniformes} (GLU), a
software package which manages a distributed dictionary of resources
(Fernique et al. \cite{glu}).
Each resource is described by its address, the query syntax, 
test information, links to
description and help files, etc.  The {\it GLU Dictionary} descriptions are
maintained up-to-date by each service provider 
and shared among all participants. The {\it GLU
Resolver} allows the service manager to use symbolic names, 
instead of physical names, for the links; these names
are then translated on the fly using the information contained in the
{\it GLU Dictionary}. 

The GLU development has allowed CDS to build reliable links between its own
services, to manage mirror copies,
and to implement a common presentation of the
CDS pages, with homogenized headers. 

Moreover, the GLU is being shared with
all the partners of the AstroBrowse NASA initiative:
information retrieval tools are being
developed for providing a homogeneous access to a large list of
resources maintained in a common GLU Dictionary (Heikkila et al.
\cite{astrobrowse}).  One of these tools, AstroGLU (Egret et al.
\cite{astroglu}), is developed by CDS. It permits us to search on-line
services such as observatory archives, databases, etc., by coordinates,
astronomical object names, astronomer names, keywords, etc. In
fact, AstroGLU is a Web interface to the GLU Dictionary.

GLU is also used by the French
{\it Centre de Donn\'ees de la Physique des Plasmas}
(CDPP)\footnote{http://cdpp.cesr.fr/}.

\subsection{The CDS role in the bibliographic network}

Starting with the {\it Bibliographical Star Index} (BSI, Ochsenbein 
\cite{bsi}) as early as
1975, CDS has always been dealing with bibliographic data: references and
objects citations in published papers are stored in {\sc Simbad}, 
and published tables in the
catalogue service. The last few years have seen a revolution
in this domain, with the extremely rapid development of electronic
publication, which has led to major conceptual evolutions in the work
of journal 
editors and publishers, and in the usage of published information by
scientists. 

The collaboration with the journal {\it Astronomy and
Astrophysics}, for which CDS implements on-line abstracts and tables
in close cooperation with the editors,
was settled in 1993, very early in the history of electronic publication
(Ochsenbein \& Lequeux \cite{aa}).
As explained in the companion paper by Ochsenbein et al. (\cite{vizier}), the
standard description of tabular catalogues proposed by CDS in 1994 has
since then been accepted by other reference journals and by the
collaborating data centers. It is now one of the important exchange
standards for astronomy, allowing for data exchange, transformation and
checks, complementary to FITS which is widely used for
binary and image data. A new standard in XML is presently being
implemented for formatting tables (Ochsenbein et al. \cite{xml}),
and to facilitate interoperability between services. In particular,
this standard has been implemented in {\sc VizieR}, and is already used for
data ingestion by {\sc Aladin}.

The CDS role in the world-wide astronomy bibliographical
network, sometimes called 
{\it Urania} (Boyce \cite{urania}), has several aspects (Lesteven
et al. \cite{lisa}):

\begin{itemize}

\item provision of selected and homogenized
bibliographic information in {\sc Simbad} and {\sc VizieR};

\item publication of `long' tables on behalf of some of the major
astronomy journals;

\item implementation of mirror copies of the {\it Astrophysical
Journal}, {\it Astronomical Journal}, and {\it Publications of the
Astronomical Society of the Pacific}, in collaboration in particular
with the
{\it University of Chicago Press}, and of NASA {\it Astrophysics Data
System} abstracts and scanned images of articles;

\item active participation to the definition of exchange standards;

\item R\&T efforts to handle `textual' information, which have
led to the development of {\it Document maps}, using the technique of
`Self-Organizing maps' (Kohonen, \cite{origk}) for displaying
references classified on the basis of the semantic proximity of their
contents (Poin\c{c}ot et al. \cite{kohonen}).

\end{itemize}

The definition of exchange standards such as the bibcode and the
standard description of tables, the close collaboration with the
journals and the ADS, have permitted an excellent synergy among
the on-line bibliographic services. For instance, data exchange,
links, exchange and installation of mirror copies, have been
implemented between CDS and ADS, which also uses {\sc Simbad} as a name
resolver. The on-line versions of {\it Astronomy and
Astrophysics} and the {\it Supplement Series} contain links to
the CDS catalogue  service, as part
of the publication, and to the list of {\sc Simbad} objects for each paper.

The Data Center has also brought new
methods to validate the journal contents, complementary to the referees'
work: tools have been developed to check the consistency of data in
electronic tables, and detected errors are reported directly to the author
by CDS before publication, and corrected. 

In addition, the development of semi-automatic methods for
recognition of astronomical object names in texts is being
studied (Lesteven at al.
\cite{lisa}). This is 
rendered difficult by the extreme complexity of astronomical
nomenclature, but there are potentially innovative applications,
such as building links between object names in journal articles and
the information contained in {\sc Simbad}. A prototype implementation is
operational at CDS in a simple case (object names in abstract
keywords). {\it New Astronomy} also provides links from object names in
articles to {\sc Simbad} and NED, with manual tagging and verification. But many
fundamental questions remain to be solved, e.g. the management of links
between object names in journals that remain unchanged, and object
names contained in databases which may change. 

\subsection{The CDS role for the access to observation archives}

The objective is to use the CDS as a `hub' to observatory archives: each
CDS service, with its own functionalities, allows the user to select the
observation he or she would like to check, and to access these
observations through an http link to the archive service.

{\sc VizieR} is potentially a major tool to access observatory databases: the
archive holdings are normally listed in a `log', i.e. in a table which
contains the list of available observations with some additional
information, such as the instrument mode, time and duration of
observation, target position, target name, PI name, etc. Data in tabular
form are very easy to include in {\sc VizieR} -- one just has to build their
description in standard format. A data archive log included in {\sc VizieR}
can be searched by querying any of its fields, thus allowing the user to
select the information of interest. The next step is to build links
between the log entries in {\sc VizieR}, and the data in the archive:
this is already operational for several archives, in collaboration with
the data providers, and using the GLU to implement the links. 
One also has to update evolving
logs, for implementing links to on-going space missions or 
ground-based programs. 
This has been developed in recent years, and is now
fully operational. In November 1999, {\sc VizieR} 
was able to access the FIRST/VLA survey data, and the IUE and 
HST archives. Discussions are under way with several
other projects.

Implementation of links from {\sc Simbad} to data archives is less
straightforward, since the logs are usually not easy to
cross-identify with the database. This is done on a
case-by-case basis. Links to IUE and HEASARC are available at
present time: the IUE log has been cross-identified with {\sc
Simbad}, taking advantage of the fact that CDS had homogenized
the mission target nomenclature on behalf of ESA (Jasniewicz et
al. {\cite{iue}); for the links to HEASARC, the high energy
objects are recognized by checking the list of identifiers for names
coming
from a high-energy mission (e.g., RX or 1RXS, among others, for
ROSAT). More will be done in the future through the implementation 
in {\sc Simbad} of
links pointing to {\sc VizieR}.

{\sc Aladin} gives access to data archives through their logs in {\sc VizieR}, and is also
able to display archive images. This is a major evolution towards a
comprehensive tool permitting comparison of images at different resolutions or
wavelengths, with active links to the original data.

\subsection{Dealing with large surveys}

The large surveys underway or planned at different wavelengths, such as
DENIS and 2MASS in the infrared, SLOAN at optical wavelengths, the large
Schmidt telescope plate catalogues (GSC I and II, USNO, APM, 
etc.), play an important role, 
both for multi-wavelength studies, and by providing reference
objects. Astronomers thus need
easy access to the data of each survey, and also tools
to use the data from one survey, together with information from other
origins. These needs have recently been summarized in the concept of
`Virtual Observatory'(see e.g.\ Szalay \& Brunner \cite{terabyte}).

CDS has been involved in active discussions with the major survey 
projects in the
last few years. As explained in Ochsenbein et al. (\cite{vizier}), an
efficient method to query very large tables by position 
has been implemented in the
CDS catalogue service, with the same user interface as {\sc VizieR},
for tables larger than the few million
objects manageable in relational systems. The USNO
catalogue (520 million objects), the public data of DENIS and 2MASS, have
been made rapidly
available in this service. The APM catalogue will also be installed
soon, as well as GSC II as soon as it will be publicly available.
{\sc Aladin} gives access to the surveys implemented in {\sc VizieR}, and is very
useful for data validation and for the assessment of criteria for
statistical cross-identification.

Moreover, CDS has been contributing to the DENIS project, by 
developing an
on-line service to distribute public and private information
(Derriere et al. \cite{deniscds}), and data comparison with the
information in the other CDS services has already served for data
validation. CDS also participates in TERAPIX (data pipeline of the CFHT
MEGAPRIME project): it will distribute the result catalogue and probably
also summary images.

In addition to the present access to very large catalogues by coordinate
queries, evaluation of the usage of commercial Object Oriented database
systems for multicriteria access to very large catalogues 
is under way (Wenger et al. \cite{oodb}). Moreover, the {\it ESO/CDS
Data Mining project} aims at accessing and combining information stored at ESO
or CDS, and to perform cross-correlations in all the
parameter space provided by the data catalogues -- not restricting the
correlations to positional ones (Ortiz et al. \cite{datamining}).

\section{Conclusion}

The usage of the CDS services has undergone a revolution in the last
few years, with the outcome of the World Wide Web, allowing easy access to
on-line information, the integration of data and documentation,
and navigation between distributed information. This means an
explosion in the usage of the services, new
functionalities, new concepts in the partnership of
data centers and journals, since published information can now be
considered as data (e.g., published tables are now usable like reference
catalogues), and an increase in international partnership, to build up
links and to define exchange standards. In parallel, the construction
of the
database contents remains a long-term activity, with increasing
volume of information to deal with, 
and high standards of scientific and technical expertise
needed in the value-added data center activities.

Navigation and links will certainly remain important keywords for the
future, and the topic of interoperability is clearly emerging. One
aspect is the construction of links between distributed services.
Another one is the building of comprehensive information retrieval tools,
as stressed in the AstroBrowse NASA initiative. To go further, one needs
to be able to integrate the result of queries to heterogeneous services
(ISAIA project, Hanisch \cite{isaia}).
In this context, the elaboration of exchange standards and
metadata descriptions common to all service providers are fundamental
keys to success. On a technical point of view,
XML may be one important tool for data integration.
{\sc Aladin} is an example of a comprehensive tool, allowing integration of
reference images, with information from catalogues, databases and data
archives.

The rapid development of the world-wide `bibliographic network' 
has been particularly
impressive, and the `data archive network' seems well under way. The CDS
is a major hub in the on-line `Virtual Observatory' presently
under construction: its services allow astronomers to select
the information of interest for their research, and to access
original data, observatory archives and results published in
journals.

\begin{acknowledgements}
CDS acknowledges the support of INSU-CNRS, the Centre National
d'Etudes Spatiales (CNES), and Universit\'e Louis Pasteur. Many
of the current developments have been made possible by
long-term support from NASA, ESA and ESO,
and {\it Astronomy \& Astrophysics}.
Many other partners are involved in building up the
international astronomy network, among which are the
AAS, ADS, HEASARC, IPAC and NED, STScI, 
ADC, CADC, INASAN, NOAJ, and many others which cannot
all be cited here.

Developing and maintaining the data-bases is a collective
undertaking. The expertise and dedicated work of the
documentalists, engineers and astronomers who work for CDS in
Strasbourg and elsewhere are the foundations of the quality of the
services. All of them are associated with this paper.
Long term support from Institut d'Astrophysique de Paris, Observatoire
de Paris (DASGAL), Observatoire de Bordeaux, GRAAL (Montpellier) 
and Observatoire Midi-Pyr\'en\'ees (Toulouse), is gratefully
acknowledged.

We thank Jean Delhaye, Jean Jung, Carlos Jaschek, and Michel
Cr\'ez\'e, for their vision and leadership in the different phases of
the CDS project.
\end{acknowledgements}


\begin{thebibliography}{}

\bibitem[2000]{aladin}
Bonnarel, F., Fernique, P., Bienaym\'e, O., et al., 2000,
A\&AS, this issue

\bibitem[1998]{urania}
Boyce, P.B., 1998, in {\it LISA III}, eds.\ U. Grothkopf,, H. Andernach, 
S. Stevens-Rayburn, M. Gomez, ASP Conf. Ser. 153, p.107

\bibitem[2000]{deniscds}
Derriere, S., Ochsenbein, F., Egret, D., 2000, in {\it ADASS IX}, ASP Conf.
Ser. in press

\bibitem[1998]{astroglu}
Egret, D., Fernique, P., Genova, F., 1998, in {\it ADASS VII}, 
eds.\ R. Albrecht, R. N. Hook, H. A. Bushouse, ASP Conf.
Ser. 145, p.416

\bibitem[1996]{mirror}
Eichhorn, G., Accomazzi, A., Grant, C. S., et al., 1996, BAAS 189,
\#06.01

\bibitem[1999]{denis}
Epchtein, N., Deul, E., Derriere, S., et al., 1999, A\&A 349, 236

\bibitem[1997]{hip}
ESA, 1997, {\it The Hipparcos and Tycho catalogue}, ESA--SP 1200

\bibitem[2000]{java}
Fernique, P., Bonnarel, F., 2000, in {\it ADASS IX}, ASP Conf. Ser. in press

\bibitem[1998]{glu} 
Fernique, P., Ochsenbein, F., Wenger, M., 1998,
in {\it ADASS VII}, eds.\ R. Albrecht, R. N. Hook, H. A. Bushouse,
ASP Conf. Ser. 145, p. 466

\bibitem[2000]{isaia}
Hanisch, R.A., 2000, in {\it ADASS IX}, ASP Conf. Ser. in press

\bibitem[1997]{starpages}
Heck, A., 1997, in {\it Electronic Publishing for Physics and Astronomy},
ed.\ A. Heck, Kluwer, p. 211

\bibitem[1999]{astrobrowse2}
Heikkila, C.W, McGlynn, T.A., White, N.E, 1999, in {\it ADASS VIII},
eds.\ D. M. Mehringer, R. L. Plante, D.
A. Roberts, ASP Conf. Ser. 172, p. 221

\bibitem[1995]{astroweb}
Jackson, R.E., Wells, D., Adorf, H.-M., et al., 1994, A\&AS
108, 235

\bibitem[1990]{iue}
Jasniewicz, G., Egret, D., Barylak, M., Wamsteker, W., 1990, in {\it
Evolution in Astrophysics: IUE Astronomy in the Era of New Space
Missions}, ESA Report, p. 601

\bibitem[2000]{ads}
Kurtz, M.J., Eichhorn, G., Accomazzi, A., et al., 2000, this issue

\bibitem[1982]{origk}
Kohonen, T., 1982, Biological Cybernetics 43, 59

\bibitem[1998]{lisa}
Lesteven, S., Bonnarel, F., Dubois, P., et al.,
1998, in {\it LISA III}, op. cit.
p.61

\bibitem[1994]{dic}
Lortet, M.C., Borde, S., Ochsenbein, F., 1994, A\&AS 107, 193

\bibitem[1997]{astrobrowse}
McGlynn, T.A., White, N.E., Fernique, P., Wenger, M, Ochsenbein, F.,
1997, BAAS 191, \#17.04

\bibitem[1982]{bsi}
Ochsenbein, F., 1982, in {\it Automated Data Retrieval in Astronomy}, eds.\
C. Jaschek \& W.D. Heintz, IAU Coll. 64, Dordrecht, D. Reidel
Publishing Company, p.171

\bibitem[1995]{aa}
Ochsenbein, F., Lequeux, J., 1995, Vistas in Astron. 39, 227

\bibitem[2000]{vizier}
Ochsenbein, F., Bauer, P., Marcout, J., 2000, A\&AS this issue

\bibitem[2000]{xml}
Ochsenbein, F., Albrecht, M., Brighton, A., et al., 2000, 
in {\it ADASS IX}, in press

\bibitem[1999]{datamining} 
Ortiz, P.F., Ochsenbein, F., Wicenec, A., 
Albrecht, M., 1999,
in {\it ADASS VIII}, eds.\ D. M. Mehringer, R. L. Plante, D.
A. Roberts, ASP Conf. Ser. 172, p. 379

\bibitem[1998]{kohonen}
Poin\c{c}ot, P., Lesteven, S., Murtagh, F., 1998, A\&AS 130, 183

\bibitem[1995]{bibcode}
Schmitz, M., Helou, G., Dubois, P., et al.,
1995, in {\it Information \& On-line Data in
Astronomy}, eds.\ D. Egret \& M. Albrecht, Kluwer Acad. Publ., p.259

\bibitem[1998]{terabyte}
Szalay, A.S., Brunner, R.J., 1998, in {\it New Horizons from
Multi-Wavelength Sky Surveys}, IAU Symp. 179, eds.\
McLean, B.J., Golombek, D.A., Hayes, J.J.E., Payne, H.E.,
Kluwer Acad. Publ., p.455

\bibitem[2000a]{simbad}
Wenger, M., Ochsenbein, F., Egret, D., et al., 
2000a, A\&AS this issue

\bibitem[2000b]{oodb}
Wenger, M., Kinnar, F., Jocqueau, R., 2000b, in {\it ADASS IX}, ASP Conf.
Ser. in press

\end{thebibliography}
\end{document}